\documentclass[reprint,preprintnumbers,amsmath,amssymb,nofootinbib,aps,prx]{revtex4-1}

\usepackage{slashed}
\usepackage{amsmath}
\usepackage{amssymb}
\usepackage{amsthm}
\usepackage{hyperref}
\usepackage{indentfirst}
\usepackage{psfrag}
\usepackage{graphicx}
\usepackage[utf8]{inputenc}

\hypersetup{colorlinks=true, linkcolor=blue, urlcolor=blue, citecolor=blue, linktocpage=true}

\begin{document}

\title{Coset space construction for the conformal group. I. Unbroken phase.}

\author{I. Kharuk}
\email{ivan.kharuk@phystech.edu}
\affiliation{Moscow Institute of Physics and Technology, \\
Institutsky lane 9, Dolgoprudny,  Moscow region, 141700, Russia,}
\affiliation{Institute for Nuclear Research of the Russian Academy of Sciences,
\\ 60th October Anniversary Prospect, 7a, Moscow, 117312, Russia}

\preprint{INR-TH-2017-025}

\begin{abstract}

The technique for constructing conformally invariant theories within the coset space construction is developed. It reproduces all consequences of the conformal invariance and Lagrangians of widely-known conformal field theories. The method of induced representations, which plays the key role in the construction, allows to reveal a special role of the ``Nambu-Goldstone fields'' for special conformal transformations. Namely, their dependence on the coordinates turns out to be fixed by the symmetries. This results in the appearance of the constraints on possible forms of Lagrangians, which ensure that discrete symmetries are indeed symmetries of the theory.

\end{abstract} 

\maketitle

\section{Introduction and summary}
\label{sec:1}

The coset space technique (CST) is a very powerful tool for obtaining Lagrangians with non-linear\footnote{More precisely, such representations are non-homogeneous. In what follows those two types of realizations would not be distinguished.} realization of symmetries. Its area of applicability covers the construction of effective Lagrangians resulting from spontaneous symmetry breaking \cite{Coleman:1969sm,Callan:1969sn,Ogievetsky1974} and of theories with unbroken yet non-linearly realized symmetries, such as Poincare or gauge invariance \cite{Ivanov:1976zq,Goon:2014ika,Ivanov:1981wn,Goon:2014paa}.

However, the question of how to apply this technique to the construction of Lagrangians with unbroken conformal invariance remains open. The problems in this route stem from the fact that, in order to make the CST applicable, one must take the corresponding coset space to be
\begin{equation} \label{ConfAlgCosetExtend}
g_H = e^{iP_\mu x^\mu} e^{iK_\nu y^\nu} \;,
\end{equation}
where $ P_\mu $ and $ K_\nu $ are generators of translations and special conformal transformations (SCT) accordingly. While the first term in the coset above is standard, the second one is not. Hence, to make sence of the theories obtained from coset (\ref{ConfAlgCosetExtend}), one should assign $ y^\nu $ a suitable interpretation. Several ideas on this point were suggested in \cite{Salam:1970qk,Wehner:2001dr,Ivanov:1981wm}. However, none of them can be claimed fully successful, since they include \textit{ad hoc} prescriptions or use methods beyond the CST.

The aim of this paper is to revisit the application of the CST to the construction of conformally invariant theories. By paying a careful attention to the discrete elements of the conformal group and to the method of induced representations \cite{Mackey:1969vt}, the correct usage of the CST in this case is obtained. It allows to reproduce all consequences of conformal invariance in a technically natural way and clarifies a special role of $ y^\nu $, the ``Nambu-Goldstone field'' for SCT. More specifically, the suggested construction is based on the fact that conformal field theories (CFT) are defined on a (pseudo--)sphere. As it is well known, its minimal atlas consists of two coordinate charts --- one at the south and one at the north poles of the sphere, each covering the whole sphere except for the opposite pole. The coordinates in these patches, $ \tilde{x}^\mu $ and $ \tilde{y}^\nu $ respectively, must obey the proper gluing map in the overlapping region,
\begin{equation} \label{GlueInt}
\tilde{y^\nu} = \frac{\tilde{x}^\nu}{\tilde{x}^2} \,, ~~ \tilde{x}^\nu \neq 0 \;,
\end{equation}
and vice versa. This results in a fact that for the construction of CFT's Lagrangians in the CST framework one should use coset space (\ref{ConfAlgCosetExtend}) in which $ x^\mu $ play the role of the coordinates and $ y^\nu $ is a field. Furthermore, the constructed Lagrangians must admit (\ref{GlueInt}) as a solution of $ y^\nu $'s  equations of motion (EqM), which imposes strong constraints on the possible forms of Lagrangians and is a qualitatively new requirement one must fulfil. The geometrical meaning of this condition is that it turns $ 2d $--dimensional coset space (\ref{ConfAlgCosetExtend}) into a $ d $--dimensional sphere, which is the space CFT are defined on. In particular, it turns out that this requirement ensures that the virial current of constructed theories is a total derivative, which is a well known property of CFT's.

The novelty of the developed technique comes in two aspects. First, the suggested approach allows obtaining homogeneously transforming quantities and conformally invariant Lagrangians directly, by the means of CST. This can be considered as a step towards in obtaining CFTs since this simplifies the analogous construction of \cite{mack1994finite}. Namely, unlike the latter work, in the developed formalism there is no need in introducing fields in an auxiliary space and then projecting them to physical states. Secondly, the developed formalism reveals the special role of the inversion, the discrete element of the conformal group, in the coset space formalism. Specifically, its proper consideration reveals the underlying geometrical construction and the role of the ``Nambu--Goldlstone'' fields for SCT. Having this understanding is important since CST is widely used for obtaining effective Lagrangians resulting from the spontaneous breakdown of the conformal invariance \cite{Ivanov:1975zq,Volkov:1973vd,Hinterbichler:2012mv}. However, until the correct application of the CST in the unbroken phase is established, one can argue such constructions to be disputable. The extension of the developed formalism to a spontaneously broken phase and connected topics will be addressed in the proceeding paper.  

The paper is organized as follows. In section \ref{sec:2}, by discussing the properties of the conformal group, the basis for the subsequent construction is formed. In section \ref{sec:3}, the technique allowing obtaining conformally invariant Lagrangians within the CST is developed. Section \ref{sec:4} is devoted to the discussion of the results, including non--mathematical interpretation of the suggested technique, and concludes the paper. In appendices \ref{AppendA} and \ref{AppendB}, a generalization of the presented technique to other spacetime groups is given. 

For clarity, in the paper the conformal group is taken to be Euclidean, reformulation of the result to the Minkowski spacetime is straightforward.

\section{The Conformal group}
\label{sec:2}

An arbitrary element of the conformal group can be presented as a product of five basic elements,\footnote{This follows from the isomorphism $ \text{Conf}(d) = O(1,d+1) $. In particular, as $ O(1,d+1) $ can be considered as a symmetry group of the hypersurface $ - y_0^2 + y_1^2 + ... + y_{d+1}^2 = 0 \subset \mathbb{R}^{1,d+1} $, this allows to give a strict definition of $ I $ as the element of $ O(1,d+1) $ changing the sign of $ y_0\, $.}
\begin{equation} \label{ConfGen}
\text{Conf}(d) = \lbrace \, e^{iP_\mu a^\mu},~e^{iL_{\mu\nu}\omega^{\mu\nu}}\,,~e^{iD\sigma}\,,~R\,,~ ~I \, \rbrace 
\end{equation}
where $ L_{\mu\nu}  $ and $ D $ are generators of the Lorentz transformations and dilations accordingly, $ R $ is the reflection of the coordinates, $ I $ is the inversion, and $ \mu = 1, .. , d \, $. In particular, the conformal group has the involute group automorphism generated by the inversion, 
\begin{equation} \label{CFTAuto}
G ~ \rightarrow ~ G: ~~~  \forall \, g \in G  ~ \rightarrow ~ I \, g \, I \;,
\end{equation}
which allows to reveal a special role of the inversion. Namely, under automorphism (\ref{CFTAuto}) the basic elements are mapped as
\begin{equation} \label{GroupAutoInv}
\begin{gathered}
I\, e^{iP_\mu a^\mu} I \equiv e^{iK_\mu a^\mu}\,, ~~~ I \, e^{iL_{\mu\nu}\omega^{\mu\nu}} I = e^{iL_{\mu\nu}\omega^{\mu\nu}}\,,  \\ ~~~ I \, e^{iD\sigma} I = e^{-iD\sigma}\,, ~~ IRI = R\;.
\end{gathered}
\end{equation}
The first relation in the formula above, in fact, defines SCT. This observation qualitatively differs the role of $ I $  from that of $ R $, since the latter invokes automorphism mapping group elements to themselves (up to a sign) only. As this makes the role of $ R $ trivial, further its presence will be ignored, while keeping track of $ I $ will be of crucial importance.

A $d$-dimensional homogeneous space of the conformal group is known to be a sphere $ S^d $, which is equivalent to the Euclidean space supplemented by a point at infinity. In particular, since $ I $ exchanges the origin and the point at infinity, the latter cannot be dropped. Note that the standard atlas of $ S^d $ consists of two charts that include the south ($S$) and north ($N$) poles of the sphere accordingly. In particular, these charts can be naturally ``created'' by acting by translations and SCT on $S$ and $N$ correspondingly. This constitutes the second observation that will be important for working out the proper application of the CST to the conformal group.

\section{Coset space construction for the Conformal group}
\label{sec:3}

\subsection{Establishing the proper coset space}
\label{sec:3-1}

Before discussing the application of the CST to the conformal group, let us remind the reader its standard rules. Let $ G $ be some symmetry group and $ \mathcal{A} $ its (chosen) homogeneous space. Then, to construct $ G $-invariant Lagrangians for the fields defined on $ \mathcal{A} \,$, one should follow the steps below \cite{Mackey:1969vt}:
\begin{itemize}
\item Define the stability group $ H $ of some point of $ \mathcal{A}\, $.
\item Introduce $ \mathcal{V}\, $ --- a space of a representation of $ H $.
\item Promote this representation to that of $ G $. This is done by redefining the elements of $ \mathcal{V} $ as functions with the domain $ \mathcal{A} $ and defining the action of $ G $ on them according to the standard CST rules.
\item Calculate the Maurer-Cartan forms (MCF) for the coset space $ G/H $ and use them for the construction of $ G $-invariant Lagrangians. 
\end{itemize} 
In this procedure, the first three steps constitute the method of induced representations, which is an intrinsic part of the CST.

However, an attempt to apply the recipe above to the conformal group and its homogeneous space $ S^d $ fails. Indeed, identifying the stability group of $ S $ and $ N $ yields,
\begin{equation} \label{ConfHomCoset}
\begin{gathered}
S: ~~~ S^d = \text{Conf}(d) / ( SG(d) \times P )\,, \\
N: ~~~ S^d = \text{Conf}(d) / ( SG(d) \times K ) \;,
\end{gathered}
\end{equation} 
where $ SG(d) = SO(d) \times D \,$, $ P = \lbrace e^{iP_\mu y^\mu } \rbrace $ is the group of translations, and $ K = \lbrace e^{iK_\nu y^\nu } \rbrace \, $. Further, to obtain MCF, one should take a logarithmic derivative of one of the coset spaces above. However, since both of them include the inversion, this step cannot be done, thus yielding the whole procedure inapplicable. At the same time, the coset space without the inversion is too small for building representations of the conformal group, since it does not cover one of the poles of the sphere. 

One may try to avoid this problem by considering representations of the group obtained from $ \text{Conf}(d) $ by excluding the inversion. In this case, the fields of a theory are defined on the Euclidean space, and the corresponding coset space reads
\begin{equation} \label{ConfAlgCoset}
g_H = e^{iP_\mu x^\mu} \;.
\end{equation}
However, such coset space is not homogeneously reductive,\footnote{By definition, a coset space $ G/H $ is homogeneously reductive if $ [Z,V] \subset Z\ $ and $ [V,V] \subset V\, $, where $ V_i $ are the generators of $ H $ and $ Z_a $ supplement them to the full set of generators of $ G $.} and thus cannot be used for the construction of conformally invariant Lagrangians \cite{Ogievetsky1974}. To fix this issue, one can extend coset space (\ref{ConfAlgCoset}) to (\ref{ConfAlgCosetExtend}) \cite{Salam:1970qk,Wehner:2001dr,Ivanov:1981wm}, but this results in the emergence of two new problems. The first one is that it is an \textit{ad hoc} prescription, and, hence, leaves the interpretation of $ y^\nu $ unclear. On the one hand, it cannot be interpreted as a field, since (unbroken) CFTs do not necessarily posses such field. And on the other hand, considering them as additional coordinates is unnatural, since the fields of a theory are defined on a $ d $-dimensional manifold. Independently from the solution of this problem, the second one unveils an inconsistency of this approach. To reveal it, one should act by SCT on (\ref{ConfAlgCosetExtend}) and read out the transformation law of the coordinates, \footnote{This formula is proven in the CST framework in section \ref{sec:3-3}. }
\begin{equation} \label{SCTCoord}
x'^\mu = \frac{x^\mu + b^\mu x^2}{1 + 2b_\mu x^\mu + b^2 x^2}\;,
\end{equation}
where $ b^\mu $ are the parameters of the applied SCT and $ x'^\mu $ are the transformed coordinates. Then, the problem is that the point for which the denominator in (\ref{SCTCoord}) is zero is mapped to the infinity, while none of the points of $ \mathbb{R}^d $ are mapped back to it. The only way to solve this problem is to consider $ S^d $ as the space the fields are defined on, and, consequently, construct representations of the conformal group.\footnote{Note that one cannot solve this problem by restricting the theory to some vicinity of the origin, since this makes the action of the translations on coset (\ref{ConfAlgCosetExtend}) ill-defined.}

To establish the way how the CST can be applied to the construction of conformally invariant theories on $ S^d $, note that the problems encountered in the previous paragraphs stem form the fact that an atlas of $ S^d $ must contain at least two coordinate charts. Indeed, for a Lie group $G$, one can introduce coordinates on $G$, as well as on its quotient spaces $ G/H $. Any homogeneous space of $G$ is isomorphic to $ G/H $, where $H$ is the stability group of some point of $ \mathcal{A} $. Then, if $ \mathcal{A} $ can be covered by one coordinate chart, the coordinates on $ \mathcal{A} $ suggested by this isomorphism are well-defined on the whole $ \mathcal{A} $. However, if it is not the cases, they become ill-defined around some points of the manifolds. For example, for the conformal group and its homogeneous space $ S^d $, the differentials of the coordinates introduced via isomorphism (\ref{ConfHomCoset}) become singular around one of the poles of the sphere. 

To resolve this problem, one can employ the following trick. Consider a coset space obtained by factorizing $ \text{Conf}(d) $ over a subgroup leaving both poles of the sphere invariant up to their exchange,
\begin{equation} \label{ConfCoset}
g_H = \text{Conf}(d) / (SO(d) \times D \times I) = e^{iP_\mu x^\mu}e^{iK_\nu y^\nu}\;.
\end{equation}
If one considers $ x^\mu $ and $ y^\nu $ as independent coordinates, then coset space (\ref{ConfCoset}), as a group manifold, is of dimension $ 2d $. However, if one requires $ x^\mu $ and $ y^\nu $ to be connected via the gluing map of the coordinate charts around the south and north poles of the sphere,
\begin{equation} \label{Glue}
y^\nu (x) = \frac{x^\nu}{x^2}\,, ~~~\vec{x} \neq \vec{0}\;, 
\end{equation}
and vice versa, (\ref{ConfCoset}) becomes isomorphic to a sphere $ S^d $, which is exactly the space the fields of CFTs are defined on. This prescription is nothing but the standard rule of gluing together two spaces into a new one used in the surgery theory of manifolds and, in particular, patchwork. Then, this suggests that one can employ the following prescription for the construction of conformally invariant Lagrangians:
\begin{itemize}
\item Start with coset space (\ref{ConfCoset}), which is homogeneously reductive, and consider $ x^\mu $ as coordinates and $ y^\nu $ as a function thereof.
\item Extract homogeneously transforming quantities from Maurer--Cartan form for coset space (\ref{ConfCoset}).
\item Construct conformally invariant Lagrangians as $ SG(d) $-invariant products thereof that admit (\ref{Glue}) as a solution.
\end{itemize} 
Importantly, this implies considering $ x^\mu $ as the only independent coordinates, and, hence, all of the fields should be introduces as a function of $ x^\mu $ only. Before implementing this procedure in practice, the following four points should be commented.

First of all, the discussion above lacks strict mathematical evidence for the suggested construction. In appendix \ref{AppendA} it is shown that the usage of coset space (\ref{ConfCoset}) and condition (\ref{Glue}) follow from geometrical considerations --- they allow us to endow $ S^d $ with an atlas structure. This forms the most fundamental and strict grounding of the suggested formalism. In section \ref{sec:4} another interpretations of the developed technique will be given, which is not mathematically strict but has a clear interpretation. Here we would also like to note that coset (\ref{ConfCoset}) can be thought of as acting on both poles of the sphere simultaneously with additional requirements that 1) $ P^\mu $ and $ K^\nu $ act non-trivially only on $ S $ and $ N $ respectively, and 2) the points of these orbits that are mapped to each other by the action of the inversion are identified. This makes it natural to dub coset (\ref{ConfCoset}) ``two-orbit'' coset space. 

Secondly, in the suggested procedure $ x^\mu $ and $ y^\nu $ are considered on different grounds: $ x^\mu $ are independent coordinates, while $ y^\nu $ are functions thereof. This prescription is justified in the next section by the method of induced representations. It should be mentioned, however, that their roles can be exchanged, since both of them provide coordinates on the whole $ S^d $ expect for one of the poles.

Further, it should be noted that condition (\ref{Glue}) may not be compatible with the symmetries, or that Lagrangians satisfying this requirement do not exist. However, in the next section it is shown that (\ref{Glue}) is not only compatible with the symmetries but is required by them. Moreover, in section \ref{sec:3-4} it is shown that all known Lagrangians of conformal field theories can be reproduced within the suggested technique.

Finally, note that in the reasoning that led to the procedure above it was not laid that coset space (\ref{ConfCoset}) will necessarily be homogeneously reductive. However, it turned out to have this property. The fundamental reason why this happened to be the case and why, as it is shown below, coset space (\ref{ConfCoset}) leads to the well-defined transformation properties of the translational MCF, is explained in appendix \ref{AppendB}.  

\subsection{Compatibility with the symmetries}
\label{sec:3-2}

To show that condition (\ref{Glue}) is, in fact, required by the conformal symmetry, consider first the following two ways of inducing (in two stages) a representation of $ SG(d) $ to that of $ \text{Conf}(d) $,
\begin{equation} \label{RepScheme}
\begin{gathered}
SG(d):(\psi)  \rightarrow \left[ \begin{array}{lr}
SG(d) \times K:  \left( y^\nu, \psi(y) \right) \\ 
SG(d) \times P:  \left( x^\mu, \psi(x) \right)
\end{array} \right.
\rightarrow  \\ \rightarrow \left[
\begin{array}{lr}
\text{Conf}(d): \left(x^\mu, y^\nu(x), \psi(x) \right) \\
\text{Conf}(d): \left(y^\nu, x^\mu(y), \psi(y) \right)
\end{array} \right. \;,
\end{gathered}
\end{equation} 
where arrows indicate an extension of a representation and given in parentheses are the elements of the space of a representation at the corresponding stage. At the final step one should also introduce the action of $ I $ as the inversion of the coordinates. Alternatively, one can induce the same representation of $ SG(d) $ to that of the conformal group directly. According to the theorem on induction in stages \cite{Mackey:1969vt,Blattner1961}, the resulting representations are equivalent. Then, by transitivity, the two representations constructed in (\ref{RepScheme}) are equivalent as well. This can be the case if and only if $ y^\nu $ and $ x^\mu $ are connected via gluing map (\ref{Glue}), which allows to switch between representations (\ref{RepScheme}) by change of coordinates (\ref{Glue}). Thus, the explicit forms of $ y^\nu(x) $ and $ x^\mu(y) $ in the upper and lower schemes of induction in (\ref{RepScheme}) accordingly are fixed to provide the gluing map of the coordinate charts. 

The induction scheme leading to coset space (\ref{ConfCoset}) is
\begin{equation} \label{RepSchemeCoset}
\begin{gathered}
SG(d):(\psi) ~ \rightarrow ~ SG(d) \times I: (\psi) ~ \rightarrow \\ \rightarrow ~ \text{Conf}(d): \left( x^\mu, y^\nu, \psi(x,y) \right)\;.
\end{gathered}
\end{equation}
Indeed, obtained in this way, the fields of the theory are defined on a space with a doubled set of coordinates.\footnote{Note that $ (x,y) $ is a set of $ 2d $ parameters, not a two different points of the sphere.} However, since the inversion is included in the intermediate step, one should factorize $ x^\mu $ and $ y^\nu $ not only over the action of $ SG(d) $ \cite{Mackey:1969vt}, but under the action of $I$ as well. To work in this formalism explicitly, one should find the proper functional measure. This is highly non--trivial mathematical task, since, as (\ref{GroupAutoInv}) demonstrates, translations and SCT are related to each other by the action of the inversion, over which action the theory should be factorized. Instead of approaching this problem directly, one can make use of the theorem on induction in stages. Namely, it guarantees that the resulting representations for schemes (\ref{RepScheme}) and (\ref{RepSchemeCoset}) are equivalent. Then, it is possible to switch from representation suggested by (\ref{RepSchemeCoset}) to the equivalent one. This allows to consider $ x^\mu $ as the only coordinates and $ y^\nu $ as a function of $ x^\mu $ whose EqM must admit (\ref{Glue}) as a solution (or vice versa). This also shows that fields can be introduced as functions of $ x^\mu $ only.

Thus, the geometrical considerations that lead to coset space (\ref{ConfHomCoset}) and condition (\ref{Glue}), and the method of induced representations on the other hand, are in full agreement with each other. In fact, this is fully expected, since these approaches employ the same underlying construction \cite{Mackey:1969vt,hermann1966lie}.
 
Another evidence showing that (\ref{Glue}) is required by the symmetries can be obtained by studying the MCF for coset space (\ref{ConfCoset}),
\begin{equation}
g_H^{-1}d g_H = iP_\mu \omega^\mu_P + iK_\nu \omega^\nu_K + iD \omega_D + iL_{\mu\nu} \omega^{\mu\nu}_L \;.
\end{equation}
Straightforward calculation yields
\begin{equation} \label{MKFormCFT}
\begin{gathered} 
\omega^\mu_P = dx^\mu\,, ~~ \omega^\nu_K = dy^\nu + 2y_\rho dx^\rho y^\nu - y^2 dx^\nu\,, \\  \omega_D = 2 y_\rho dx^\rho\,, ~~ \omega^{\mu\nu}_L = -2 y^\mu dx^\nu \;.
\end{gathered}
\end{equation} 
Since coset space (\ref{ConfCoset}) is homogeneously reductive, the action of all continuous group elements on the MCF above is well-defined, and, importantly, none of them mixes the MCF $ \omega_P^\mu $ and $ \omega_K^\nu $ with each other. However, one should also investigate the transformational properties of the MCF under the action of the inversion. The action of the latter on coset space (\ref{ConfCoset}) is equivalent to making group automorphism (\ref{CFTAuto}), thus yielding
\begin{equation} \label{InvMKTransf}
\omega_P^\mu \rightarrow \omega_K^\mu\,, ~ \omega_K^\nu \rightarrow \omega_P^\nu\,, ~ \omega_D \rightarrow - \omega_D\,, ~ \omega_L^{\mu\nu} \rightarrow \omega_L^{\mu\nu}\;.
\end{equation}
As one can see, this interchanges the 1-forms for translations and SCT. Since the inversion is a symmetry, the Lagrangians constructed within the CST must be invariant under such transformation. To understand the consequences of this requirement, note that group automorphism (\ref{CFTAuto}) also invokes the following isomorphism of $ S^d $ to itself,
\begin{equation} \label{HomAuto}
S^d ~ \rightarrow ~ S^d: ~~ \forall \, s \in S^d ~ \rightarrow ~ \hat{I}\,s \;.
\end{equation}
This mapping exchanges the coordinate charts around the south and north poles of the sphere, and, since $ x^\mu $ and $ y^\nu $ are defined as coordinates therein, leads to the exchange of their roles. Then, the equivalent way to obtain the transformed Lagrangian is to take the same exterior product of the MCF but for the coset space
\begin{equation} \label{ConfAltCoset}
\tilde{g}_H = e^{iK_\nu y^\nu} e^{iP_\mu x^\mu}\;,
\end{equation}
and with $ \omega_P^\mu $ and $ \omega_K^\nu $ exchanged. The transformed and initial Lagrangians must coincide, which is possible only if the new translational MCF, $ \omega_K^\nu = d y^\nu $, are the pullbacks of the old ones, $ \omega_P^\mu = dx^\mu $, after change of coordinates (\ref{HomAuto}). This forces $ y^\nu(x) $ to obey gluing map (\ref{Glue}), which reproduces the result obtained by the method of induced representations. 

The discussion above demonstrates that condition (\ref{Glue}), which was introduced as a way of reducing the dimensionality of coset (\ref{ConfCoset}), is, in fact, required by the symmetries. Thus, the only allowed combinations of the MCF are those that admit (\ref{Glue}) as a solution of $ y^\nu $'s EqM. This is a qualitatively new requirement one encounters in the process of applying the CST to the construction of theories on the manifolds whose atlas must contain more than one coordinate chart. 

\subsection{Reproducing representations of the Conformal group}
\label{sec:3-3} 
 
This is a convenient point to verify that the suggested usage of the CST correctly reproduces representations of the conformal group. As it follows from (\ref{RepSchemeCoset}) and subsequent discussion, fields $ \psi $ are introduced as functions of $ x^\mu $ belonging to an (irreducible) representations of $ SG(d) $ group. Hence, they are charaterized by spin $ s $ and scaling dimension $ \Delta_\psi $, which is in agreement with the common lore. To define how they transform under the action of the conformal group, one should pick arbitrary element of the conformal group, $ g \in \text{Conf}(d) $, and bring the product of $ g $ and $ g_H $ to the standard form \cite{Ogievetsky1974},
\begin{equation} \label{SCTAction}
g g_H = e^{iP x'(x,g) } e^{iK y'(x,g)} e^{iD\sigma (x,g)} e^{iL\omega(x,g)} \;. 
\end{equation} 
Note that because of the commutation relations of the conformal algebra the parameters appearing on the r.h.s. of equation (\ref{SCTAction}) are functions of $ x^\mu $ and $ g $, but not of $ y^\nu $. Then, (\ref{SCTAction}) implies that under the action of $ g ~ \psi $ and $ x^\mu $ transforms as
\begin{align} \label{MattTransfSCT}
\psi(x) \rightarrow Rep( e^{-iD\sigma(x,g^{-1})},~ &e^{-iL_{\mu\nu}\omega^{\mu\nu}(x,g^{-1})} )\psi(x)\,,  \\ \label{CoordTransSCT}
x^\mu \, \rightarrow & \; x'^\mu (x, g^{-1}) \;. 
\end{align}
where $ Rep(\cdot) $ is a representation of $ SG(d) $ appropriate for $ \psi $. It is straightforward to verify that from these rules follow the expected transformation properties of $ x^\mu $ and $\psi$ under the action of a dilataion and Lorentz transformation --- the appearing in (\ref{SCTAction}) $ \sigma $ and $ \omega^{\mu\nu} $ do not depend on the coordinates, $ x^\mu $ is a vector with scaling dimension $ -1 $ and $ \psi $ is a spin--s filed with scaling dimension $ \Delta_\psi $. To find the action of the SCT, one should act by $ g = e^{iK_\nu b^\nu} $, where $ b^\nu $ is a free parameter, on coset (\ref{ConfCoset}). This leads to the following infinitesimal versions of $ \sigma $, $ \omega^{\mu\nu} $ and $ x'^\mu $:
\begin{equation} \label{InfinitSCT}
\begin{gathered}
\sigma = 2b_\mu x^\mu \,, ~~ \omega^{\mu\nu} = b^\mu x^\nu - b^\nu x^\mu \,, \\ x'_\mu = x_\mu + 2 b_\nu x^\nu x_\mu -x^2 b_\mu \;,
\end{gathered}
\end{equation}
which coincide with the standard expressions. Then, the group property guarantees that they will coincide at the non-linear level as well. By substituting them into (\ref{MattTransfSCT}) and (\ref{CoordTransSCT}), one sees that the suggested usage of the CST correctly reproduces representations of the conformal group. In particular, note that fields $ \psi $, introduced in this way, are nothing but the so--known quasi--primary fields. Indeed, 1) they belong to irreducible representations of $ SG(d) $ group (and, hence, are characterized by spin and scaling dimension) and 2) $ \hat{K}_\mu \psi (0) = 0 $, as it follows from (\ref{MattTransfSCT}) and (\ref{InfinitSCT}). Thus, by definition, $ \psi $ are quasiprimary fields.

\subsection{Constructing conformally invariant Lagrangians}
\label{sec:3-4}

Now everything is prepared for the construction of conformally invariant Lagrangians in the coset space framework. The consideration will be restricted to the case when fields enter Lagrangian quadratically and with no more than one derivative per field,\footnote{The inclusion of higher derivative terms into consideration is non-trivial and will be carried out in a separate paper.} which is enough for the purposes of the paper. Let $ \psi $ be a field belonging to some representation of the $ SG(d) $. The 1-form associated with $ \psi $ reads \cite{Ogievetsky1974},
\begin{equation} \label{CovarDerMatterFields}
D\psi = \partial_\mu\psi dx^\mu + 2y^\nu ( \eta_{\mu\nu}\Delta + i \hat{L}_{\mu\nu})\psi dx^\mu \;,
\end{equation}
where $ \Delta $ and $ \hat{L}_{\mu\nu} $ are representations of $ D $ and $ L_{\mu\nu} $ appropriate for $ \psi $. conformally invariant Lagrangians are then obtained as $ SG(d) $-invariant wedge products\footnote{Practically, a more convenient way of obtaining conformally invariant Lagrangians is to read out the effective metric and the covariant derivatives of fields from (\ref{MKFormCFT}) and (\ref{CovarDerMatterFields}). For the details of this procedure, see, for example, \cite{Ogievetsky1974,Goon:2014ika}.}  of $ D \psi,~ \psi,~ \omega_P^\mu$, and $ \omega_K^\nu $ admitting (\ref{Glue}) as the solution of $ y^\nu $'s EqM.

The construction of conformally invariant theories will proceed from the simplest case to the most general one in three steps. Also, it will be assumed that $ d \geq 2 $. 

First, consider the case when there are no matter fields. Then, the corresponding Lagrangians describe $ y^\nu $'s ``kinetic term'',
\begin{equation} \label{SCTKinetic}
\mathcal{L}_y = \mathcal{L}_{kin}(\omega_P^\mu, \omega_K^\nu)\;,
\end{equation}
where $ \mathcal{L}_{kin}(\omega_P^\mu, \omega_K^\nu) $ is an arbitrary function constructed as a $ SG(d) $-invariant wedge product of $ \omega_K^\nu $ and $ \omega_P^\mu $. Then, since the scaling dimension of $ \omega_K^\nu $ equals one, in $ d > 2 $ the variation of (\ref{SCTKinetic}) with respect to $ y^\nu $ would always be proportional to $ \omega^\nu_K $. Hence, such theories admit the following solutions,
\begin{equation} \label{MCFConstr}
\omega_K^\nu = 0 ~~~ \Rightarrow ~~~ 
y^\nu = 0 ~~  \cup ~~ y^\nu = \frac{x^\nu}{x^2}\;.
\end{equation}
For $ d=2 $ the only possible $ SG(d) $-invariant combination of the MCF is a full derivative,
\begin{equation}
\varepsilon_{\mu\nu} \, \omega_P^\mu \wedge \omega_K^\nu = \partial_\mu y^\mu \, dx^1 \wedge dx^2 \;,
\end{equation} 
where $ \varepsilon_{\mu\nu} $ is the Levi-Civita symbol, and, hence, does not constrain $ y^\nu $'s dynamics. Thus, the requirement for $ y^\nu $ to obey the gluing map is fulfilled in this simplest case. Since $ y^\nu $'s ``kinetic term'' always admits (\ref{Glue}) as a solution, in the rest of the paper it will be omitted.

In particular, the fact that gluing map (\ref{Glue}) is a solution of the system of differential equations $ \omega_K^\nu = 0 $ can be proved on symmetry grounds. Namely, these equations are conformally invariant, and gluing map (\ref{Glue}) and $ y^\nu = 0 $ are the only functions of $ x^\mu $ having the same properties. Hence, the solutions of these equations cannot but be given by (\ref{MCFConstr}).

As the second step, consider the case when fields $ \psi_a $ mix with $ y^\nu $ only via their covariant derivatives. Let $ \mathcal{L}_\psi $ be a Lagrangian governing the dynamics of the fields. Then, varying the action with respect to $ y^\rho $, after trivial transformations, yields,
\begin{equation} \label{VarySCT}
2\frac{\delta L}{\delta D_\mu \psi_a} ( \Delta \eta_{\mu\rho} + i \hat{S}_{\mu\rho} ) \psi_a \equiv V_\rho = 0 \;,
\end{equation}
where $ V_\rho $ is the ``extended'' virial current. It includes the usual one, $ V^{(0)}_\rho $, and the term proportional to $ y^\nu $,
\begin{equation} \label{VirialCurrent}
V_\rho \equiv V^{(0)}_\rho + V^{(1)}_{\rho\nu} y^\nu \;.
\end{equation}
Note that matter fields, which have non--trivial dynamics, enter equation (\ref{VarySCT}), while there are no $ y^\nu $'s derivatives. Hence, (\ref{VarySCT}) imposes constraints on the structure of the theory, rather than on $ y^\nu $'s dynamics. Namely, (\ref{VarySCT}) shows that the extended virial current of the theory must vanish. Note that if $ V^{(0)}_\rho $ is zero, then so is full $ V_\rho $. Indeed, if $ V^{(0)}_\rho = 0 $, then the tensor structure of the first multiplier in (\ref{VarySCT}) is such that the whole expression vanishes independently from the explicit form of the former. Then, because of the latter property, the whole $ V_\rho $ vanishes as well. Thus, the suggested technique reproduces the well-known property of CFT that their virial current is identically zero. Note also that equation (\ref{VarySCT}), in fact, ensures that $ y^\nu $ disappears from the Lagrangian. This observation, combined with the study of the last case below, will allow us to suggest another interpretation of the suggested technique, which will be discussed in section \ref{sec:4}.

\textbf{Example.} A vector field theory in $ d=4 $ provides an instructive illustration of the developed technique. According to (\ref{CovarDerMatterFields}), in conformal field theories the covariant derivative of a vector field reads
\begin{equation}
D_\mu A_\nu = \partial_\mu A_\nu + 2y^\rho ( \delta_{\mu\rho} \delta_\nu^\lambda + i(\hat{S}^{(1)}_{\mu\rho})^\lambda_\nu ) A_\lambda \;,
\end{equation} 
where $ \hat{S}^{(1)}_{\mu\rho} $ is spin-1 representation of the Lorentz group. Then, the most general quadratic $ SG(d) $-invariant Lagrangian one can write is
\begin{equation} \label{GenLagrVector}
\mathcal{L} = \frac{1}{2} C^{\mu\nu\lambda\rho} D_\mu A_\nu D_\lambda A_\rho \;,
\end{equation}
where $ C^{\mu\nu\lambda\rho} $ is a constant tensor constructed from various combinations of $ \delta^{\mu\nu} $ and is symmetric in $ (\mu \nu) \leftrightarrow (\lambda \rho) $. Further, the requirement for the virial current to vanish yields 
\begin{equation}
C^{\mu\nu\lambda\rho} D_\lambda A_\rho ( \delta_{\mu\sigma}A_\nu - \delta_{\mu\nu}A_{\sigma} + \delta_{\sigma\nu}A_\mu ) = 0 \;,
\end{equation}
following from (\ref{VirialCurrent}). It can be fulfilled if and only if $ C^{\mu\nu\lambda\rho} $ is antisymmetric in its first two indices. Thus, as it was expected, the conformal invariance requires Lagrangian (\ref{GenLagrVector}) to coincide with Maxwell's one. Remarkably, it also turned out to be gauge-invariant. 

As it can be verified, the developed technique also allows to reproduce free massless spin-0 and spin-$\frac{1}{2}$ field theories in $2$ and in an arbitrary number of dimensions accordingly. Note that in case $ d \neq 2 $ the virial current of spin-0 theory is not vanishing but is the divergence of some other tensor. How one can reproduce such theories, including the so-known elastic vector field theory \cite{ElShowk:2011gz}, is explained below.

Finally, the most general class of conformally invariant Lagrangians is obtained by allowing matter fields to mix with $ \omega_K^\nu $ directly. In this case, unless the interaction terms sum up to a total derivative, the solution of $ y^\nu $'s EqM cannot be fixed to (\ref{Glue}). Consequently, one must study the question of when the interaction terms do sum up to a total derivative. This is possible only if the virial current is a divergence of some other tensor, since otherwise the linear in $ y^\nu $ term cannot be completed to a total derivative. As a straightforward but lengthy calculation in appendix \ref{AppendC} demonstrates, this is also a sufficient condition. That is, if the virial current is a total derivative, 
\begin{equation}
V^{(0)}_\rho = \partial_\mu L^\mu_\rho \;,
\end{equation}
the following Lagrangian contains the interaction terms only via full derivative,
\begin{equation} \label{OnlyViaDer}
 ~~~~ \mathcal{L} = \frac{1}{2}  D\psi \wedge \star D\psi + \varepsilon_{\mu_0 ... \mu_d} L^{\mu_0}_\nu \omega_K^\nu \wedge \omega_P^{\mu_1} \wedge ... \wedge \omega_P^{\mu_d} \;,
\end{equation} 
where $ \star $ is the Hodge dual operator. Thus, a class of scale-invariant theories, which are, in fact, conformally invariant after an improvement of the energy-momentum tensor \cite{Callan:1970ze}, correspond to the Lagrangians of type (\ref{OnlyViaDer}). 

Since the interaction terms sum up to a total derivative, Lagrangian (\ref{OnlyViaDer}) can be split in two parts,
\begin{equation} \label{WessZumino}
\mathcal{L} = \frac{1}{2} d\psi \wedge \star d \psi + d\tilde{\mathcal{L}}(y,\psi)\;.
\end{equation}
The second term in the expression above is a total derivative and, consequently, can be dropped without affecting the dynamics of the theory. Then, the first term alone can be considered as a special type of the Wess-Zumino term that can arise on the manifolds whose atlas must contain more than one coordinate chart. In particular, the standard Lagrangians for the massless spin-0 and elastic vector field theories \cite{ElShowk:2011gz} represent examples of such terms. Namely, they are obtained by dropping the corresponding total derivative part from their complete Lagrangians, which are of the form (\ref{OnlyViaDer}). 

\textbf{Example.} It is convenient to illustrate this class of theories on the process of reconstructing the $ \varphi^4 $ theory in $ d=4 $. Instead of working with the differential forms, it will be convenient to switch to the covariant derivatives of $ \varphi $ and $ y^\nu $, which can be read out from (\ref{CovarDerMatterFields}) and (\ref{MKFormCFT}) to be
\begin{equation}
D_\mu \varphi = \partial_\mu \phi + 2y_\mu \varphi \,, ~~ D_\mu y^\nu = \partial_\mu y^\nu +2 y_\mu y^\nu - y^2 \delta^\nu_\mu \;.
\end{equation}
Then, as the starting point, consider the Lagrangian
\begin{equation} \label{Phi4Start}
\mathcal{L} = \frac{1}{2}  D_\mu \varphi D^\mu \varphi + \frac{\lambda}{4} \varphi^4 \;,
\end{equation}
which is $ SG(d) $-invariant and reproduces $ \varphi $'s kinetic and potential terms. However, since (\ref{Phi4Start}) explicitly depends on $ y^\nu $, it does not admit (\ref{Glue}) as a solution and, hence, is not a valid Lagrangian. To understand whether it can be improved to include $ y^\nu $ only via full derivative, one should find the virial current of the theory, which reads
\begin{equation}
V_\rho^{(0)} = \partial_\mu \delta^\mu_\rho \varphi^2 \;.
\end{equation}
Since it is a total derivative, one can complete Lagrangian (\ref{Phi4Start}) to be of the form (\ref{OnlyViaDer}). For the case under consideration, this leads to the Lagrangian
\begin{equation} \label{Phi4Full}
\begin{aligned}
\mathcal{L} = \frac{1}{2}  D_\mu \varphi D^\mu \varphi + &\varphi^2 D_\mu y^\mu + \frac{\lambda}{4} \varphi^4 = \\ = \frac{1}{2} ( \partial_\mu \varphi )^2 &+ \frac{\lambda}{4} \varphi^4 + \partial_\mu ( y^\mu \varphi^2 ) \;,
\end{aligned}
\end{equation}
where the second line was rewritten in the form similar to (\ref{WessZumino}). In particular, the first two terms in (\ref{Phi4Full}) reproduce the standard $ \varphi^4 $'s theory. 

Summing up, the developed technique reproduces all consequences of conformal invariance and Lagrangians of the widely-known CFTs. Also, it clarifies the special role of the Nambu-Goldstone field for SCT, which is the following. The standard CST prescriptions ensure the invariance of the constructed Lagrangian under the action of the conformal algebra, while it is the condition for $ y^\nu $ to obey the gluing map that guarantees that the inversion is a symmetry of the theory as well. In particular, if the latter requirement is fulfilled, the covariant derivatives of the matter fields simplify to the usual ones on the Lagrangian level (up to a total derivative term), which explains why scale and conformally invariant Lagrangians look the same.

\section{Conclusion}
\label{sec:4}

Initially, the problems with applying the CST to the conformal group were stemming from the fact that the coset space $ \text{Conf}(d) / e^{iP_\mu x^\mu } $ is not homogeneously reductive. However, as it was demonstrated, a careful treatment of discrete symmetries allows to establish the correct way of obtaining conformally invariant Lagrangians within the CST. In appendix \ref{AppendB} this result is generalized to other spacetime groups. Namely, it is shown that the CST is applicable to groups whose homogeneous space is homogeneously reductive after the exclusion of all discrete and composite symmetries, like $ I $ and SCT in the conformal group.

As it was shown, for the construction of conformally invariant theories one should use coset space (\ref{ConfAlgCosetExtend}), as it was suggested in earlier works. Importantly, in the process of obtaining CFTs, $ y^\nu $ should be considered as a field, and constructed Lagrangians must admit gluing map (\ref{Glue}) as a solution of $ y^{\nu} $'s equations of motion. As a consequence, $ y^\nu $ must enter Lagrangians only via full derivative. This allows us to suggest the following reinterpretation of the developed technique. First, one starts with coset space (\ref{ConfAlgCosetExtend}), in which $ y^\nu $ is considered as an auxiliary field ensuring the applicability of CST. Further, to obtain theories including only physical fields, one searches for Lagrangians containing $ y^\nu $ only via full derivative. Since this requirement coincides with the one obtained in the previous section, this makes the constructions equivalent. Although this simpler approach provides a solution to the initial problem, it can be considered only as a trick, while the strict grounding of the approach follows from a careful study of the connection between the method of induced representations and CST.

To summarize the results, in the paper the method of applying the CST to the construction of conformally invariant Lagrangians was developed. A careful handling of discrete symmetries and of the geometrical meaning of the method of induced representations were found to be the keys to establishing the correct application of the CST in this case. In particular, the suggested approach reproduces the results of \cite{mack1994finite} --- conformally invariant theories are dilaton invariant and their virial current vanishes (or is a total derivative \cite{Callan:1970ze}). In \cite{mack1994finite}, this was established by studying the divergence of the conformal currents, while in the present paper it was shown how these restrictions arise from the CST perspective. Finally, the developed formalism provides a tool for the systematical construction of conformally invariant theories purely within the CST, which simplifies the standard procedure, as well as will be of use for the construction of Lagrangians with complicated symmetries, such as of the conformal-affine gravity \cite{Borisov:1974bn}.

\section*{Acknowledgements} 

The author is thankful to E. Ivanov, S. Sibiryakov and A. Shkerin for useful discussions and comments on the draft of the paper. The work was supported by the Grant 14-22-00161 of the Russian Science Foundation.

\appendix

\section{Introducing atlas structure on homogeneous spaces}
\label{AppendA}

In this appendix it is discussed which coset space should be used within the CST applied to an arbitrary Lie group $G$ and its $d$-dimensional homogeneous space $ \mathcal{A} $. By studying the geometry of such spaces, it is shown that if $ \mathcal{A} $'s atlas must contain more than one coordinate chart, this requires non-standard CST prescriptions. In particular, for the conformal group this leads to coset (\ref{ConfCoset}) used in section \ref{sec:3}.

Let $ H_0 $ be the stability group of a point $ z_0 \in \mathcal{A} $. Then, there is the isomorphism
\begin{equation} \label{IsomGen}
\mathcal{A}=G/ H_0\;.
\end{equation} 
Within this isomorphism, an element $ g_{H_0} \in G/H_0 $ is identified with the point of $ \mathcal{A} $ obtained by the action of the former on $ z_0 $. In general case, $ G/H_0 $ consists of continuous and discrete elements, which will be denoted as $ e^{iP^{(0)}_\mu x^\mu} $ and $ T_m, ~m=1,...n\,, $ accordingly (further it is assumed that there is at least one $ T_m $). Then, discrete elements are identified with a finite set of points $ \lbrace z_m \rbrace $, while $ e^{iP^{(0)}_\mu x^\mu} $ is isomorphic to $ \mathcal{A} \setminus \lbrace z_m \rbrace $. In particular, this makes it natural to refer to $ P_\mu^{(0)} $ as generators of translations and to $ x^\mu $ as coordinates on $ \mathcal{A} $. Further, for an arbitrary point $ a \in \mathcal{A}\,,~ a \notin \lbrace z_m \rbrace  $, one has
\begin{equation} \label{OrdinaryPoint}
a = e^{iP_\mu^{(0)} c^\mu} z_0 \;,
\end{equation}
where the exponential of translations is considered as an operator. In particular, on can take the differential of both sides of this formula, which demonstrates that the coordinates on $ \mathcal{A} \setminus \lbrace z_m \rbrace $ are well-defined. On the other hand, the analogy of (\ref{OrdinaryPoint}) for $ z_m $ reads
\begin{equation}
z_m = \hat{T}_m z_o \;.
\end{equation}
Unlike the previous case, one cannot consider an infinitesimal displacement of this point, since the differential of a discrete element is not defined. This demonstrates that $ G/H_0 $ cannot be covered by one set of well-defined coordinates.

The stability group $ H_m $ of a point $ z_m $ is
\begin{equation}
H_m = \hat{T}_m H_0 \hat{T}_m^{-1} \;.
\end{equation} 
Then, $ \mathcal{A} $ can also be considered as a quotient space $ G/H_m $. By repeating the reasoning of the previous paragraph, one sees that in this case the coordinates are ill-defined at all $ z_k\,, k=0,..n, $ except for $ z_m $, which is the origin of the coordinate chart. Since $ z_m $ can be chosen arbitrary, this shows that it is possible to introduce coordinates around each of $ z_k $, but they cannot be successfully extended to the whole $ \mathcal{A} $. 

The above describes a manifold whose atlas must contain at least $ n+1 $ coordinate charts. Remember that such manifolds can be obtained by considering $ n+1 $ independent coordinate charts and then gluing them together by introducing $ (n+1)! $ equivalence relations, which correspond to the gluing map of these coordinate charts. This suggest that the coset space characterizing $ \mathcal{A} $ as a manifold can be introduced as follows. All of the coset spaces $ g_{H_k} = G/H_k $, with excluded discrete symmetries, give rise to the coordinates around $ z_k $. Then, one can introduce $ n+1 $ independent coordinate charts by considering the action of the product of all $ g_{H_k} $ on $ \lbrace z_k \rbrace $ with an additional requirement that each translational generator $ P_\mu^{(k)} $ acts non-trivially only on the point $ z_k $. If $ \tilde{H} $ is the stability group of all $ z_k $ up to their exchange, and assuming that $ T_m $ mixes $ z_k $ only between each other, this is equivalent to considering the coset space $ G/\tilde{H} $,
\begin{equation} \label{GenGeomCoset}
g_{\tilde{H}} = e^{iP_\mu^{(0)}x^\mu_{(0)}} e^{iP_\mu^{(1)} x^\mu_{(1)}} ... e^{iP_\mu^{(n)} x^\mu_{(n)}} \;,
\end{equation}
where $ e^{iP_\mu^{(m)} x^\mu_{(m)}} = \hat{T}_m e^{iP_\mu^{(0)} x^\mu_{(0)}} \hat{T}_m^{-1} $, which acts on $ \lbrace z_k \rbrace $ as defined above. Obtained in this way, coset (\ref{GenGeomCoset}) describes a manifold of dimension $ d \times (n+1) $. Further, by introducing the equivalence relations 
\begin{equation} \label{GenGeomGlue}
x^\mu_{(m)} = \hat{T}_m x^\mu_{(0)} ~~~ \text{for all} ~m\;,
\end{equation} 
one glues these $ n+1 $ areas together, thus defining the atlas structure on $ G/\tilde{H} $ and making it equivalent to $ \mathcal{A} $, which is of dimension $ d $. Because $ \hat{T}_m $ form a representation of $G$, (\ref{GenGeomGlue}) defines all $ (n+1)! $ gluing maps between the coordinates charts. Moreover, because of the same property, (\ref{GenGeomGlue}) is not only automatically in agreement with the action of $ \hat{T}_m $ on coset space (\ref{GenGeomCoset}), but is also required by it. In particular, this is the reason why in section \ref{sec:3-2} the study of the MCF led to requirement (\ref{Glue}). 

Summing up the construction above, it can be said that the defining property of the coset space to be used within the CST is that it must endow the manifold under consideration with an atlas. Coset space (\ref{GenGeomCoset}) can be thought of as acting on all points $ z_m $ simultaneously, but the points of $ z_m $'s orbits must be factorized by equivalence relations (\ref{GenGeomGlue}). In particular, for the conformal group this leads to coset space (\ref{ConfCoset}) and requirement (\ref{Glue}).

\section{Reducibility of the coset}
\label{AppendB}

Taking the general set-up introduced in appendix \ref{AppendA}, denote by $ G_c $ a subset of $ G $ obtained by excluding all discrete and composite elements from the latter. For example, such procedure corresponds to excluding the inversion and SCT from the conformal group. Further, assume that: 1) $ G_c $ forms a group, 2) $ P_\mu \in G_c \, $, and 3) the algebra of $ G_c $, $ AG_c $, is homogeneously reductive with respect to the decomposition 
\begin{equation} \label{AlgDivis}
AG_c = P_\mu \oplus H_a \;,
\end{equation}
where $ H_a $ supplement $ P_\mu $ to the full set of generators of $ G_c \, $. As it will become clear shortly, such requirements are rather general. Then, the aim of this appendix is to show that the CST is applicable for the construction of $ G $-invariant Lagrangians for the fields $ \psi(x) $ defined on $ \mathcal{A} $. 

For this purpose, two technical statements need to be proved. The first one is that the action of $ H = \lbrace e^{iH_ab^a} \rbrace $ leaves not only $ \vec{0} $ invariant, but all $ z_m $ as well. Indeed, suppose otherwise --- at least for one $ k $ the action of $ H $ on $ z_k $ is non-trivial. Then, as $ H $ is a continuous group, it is legitimate to consider an infinitesimal transformation, $ e^{iH_a b^a} $. Taking $ b^a $ small enough, it can be assured that under the action of $ e^{iH_a b^a} \, z_k $ is mapped to some point $ \vec{a} \in  \mathcal{A} \setminus \lbrace z_m \rbrace $. Since the latter can be obtained by acting by $ e^{iP_\mu c^\mu} $ on $ \vec{0} $ for some $ c^\mu \, $, one has
\begin{equation}
( e^{-iH_a b^a} e^{iP_\mu c^\mu } e^{iH_a b^a} ) e^{-iH_a b^a} \vec{0} = e^{iP_\mu \tilde{c}^\mu} \vec{0}  = z_k \;,
\end{equation}
where it was used that $ AG_c $ is homogeneously reductive and that $ H $ is a stability group of $ \vec{0} $. Thus, $ z_k $ can be obtained by the action of $ e^{iP_\mu \tilde{c}^\mu} \in G_c $ on $ \vec{0} $, which contradicts the condition that $ z_k $ is identified with $ T_k $ within isomorphism (\ref{IsomGen}). This finishes the proof. Similarly, it can be proved that $ T_m $ mixes $ \lbrace z_m \rbrace $ and $ \vec{0} $ only between each other. 

The second statement is that the group automorphisms 
\begin{equation} \label{AutoGroup}
W_m:~~ G \rightarrow G\,, ~~ \forall g \in G \rightarrow T_m\, g\, T_m^{-1}\;,
\end{equation}  
map $ H $ to itself. Indeed, for a given $ m $, automorphism (\ref{AutoGroup}) can be considered as the following isomorphism of $ \mathcal{A} $ to itself,
\begin{equation} \label{IsoHomSpace}
\mathcal{A} \rightarrow \mathcal{A}: ~~~ \forall\, \vec{a} \in  \mathcal{A} \rightarrow \hat{T}_m \vec{a} \;,
\end{equation}
where $ \hat{T}_m $ is the representation of $ T_m $ acting on $ \mathcal{A} $. As it follows from the previous paragraph, (\ref{IsoHomSpace}) mixes $ \lbrace z_m \rbrace $ and $ z_0 $ only between each other. Then, since $ H $ is the stability group of  $ \lbrace z_0\, z_m \rbrace  $, isomorphism (\ref{AutoGroup}) maps it to itself, QED. 

An immediate corollary of these results is that the full set of generators of $ G $ is
\begin{equation} \label{Algebra}
P_\mu \, , ~~~ K^{(m)}_\mu \equiv T_m \, P_\mu \, T_m^{-1} \,, ~~~ H_a\;.
\end{equation}
Moreover, as $ AG_c $ is homogeneously reductive with respect to decomposition (\ref{AlgDivis}), one also has
\begin{equation} \label{ReductExtend}
[K^{(m)}_\mu , H] = T_m [P_\mu, H ] T^{-1}_m \subset K^{(m)}_\mu \;.
\end{equation}
Thus, the presence of automorphisms (\ref{AutoGroup}) strongly fixes the algebra of such groups. In particular, (\ref{Algebra}) and (\ref{ReductExtend}) correctly reproduce the structure and commutation relations of the conformal algebra.

As it was explained in appendix \ref{AppendA}, for the construction of $G$-invariant Lagrangians one should employ coset space (\ref{GenGeomCoset}). Then, as it follows from (\ref{ReductExtend}), this coset space is homogeneously reductive and, moreover, the MCF $ \omega_P^\mu $ and $ \omega_{K^{(m)}}^\mu $ do not mix with each other under the action of $ H $. Hence, one can apply the CST to ``(n+1)-orbit'' coset space (\ref{GenGeomCoset}).

In particular, note that from (\ref{ReductExtend}) it follows that $ K_\mu^{(m)} $ cannot be included into the Cartan algebra of $ G $. Hence, in such theories the fields are introduced as representations of $ \tilde{H} $, which is in agreement with the CST rules. The action of $ e^{K^{(m)}_\mu b^\mu} $ on the coordinates and fields is then given by the analogue of formulas (\ref{MattTransfSCT}) and (\ref{CoordTransSCT}) for the left action of $ G $ on $ G/\tilde{H} $. Importantly, the resulting transformations will depend on $ x^\mu $ and $ b^\mu $, but not on $ y^\mu_{(m)} $, as guaranteed by (\ref{ReductExtend}).

Note also that, because of the commutation relations (\ref{ReductExtend}), there exist $ (m+1)! $ ways of inducing (in $m$ stages) a representation of $ \tilde{H} $ to that of $G$. Then, in the same way as for the conformal group, it can be shown that $ x^{\mu}_{(m)} $s should be considered as functions of $ x^\mu $, whose explicit form must be given by (\ref{GenGeomGlue}).  

Thus, the CST can indeed be applied to coset space (\ref{GenGeomCoset}), provided that the assumptions made in the beginning of this appendix hold. The peculiarity of using CST in such cases is that the constructed Lagrangians must not only be $ \tilde{H} $-invariant combinations of the MCF for coset space (\ref{GenGeomCoset}), but admit (\ref{GenGeomGlue}) as a solution of $ x^\mu_{(m)} $'s EqM as well. This requirement complements the standard CST rules and ensures that the discrete symmetries are indeed symmetries of the theory. 

\section{Proof that the interaction terms sum up to a total derivative}
\label{AppendC}

To prove that Lagrangian (\ref{OnlyViaDer}) contains the interaction terms only via full derivative, one can take $ \psi $ to be a vector field. This does not result in the loss of generality since the integer higher order spins are formed as tensor products of spin-1 representations, while half-integer spins, except for $ 1 / 2 $, are not of physical interest. Then, assuming quadratic kinetic term, in general case it reads
\begin{widetext}
\begin{equation} \label{Kinetic}
\frac{1}{2}C^{\mu a \nu b}(D_\mu \psi_a)(D_\nu \psi_b) \equiv
\frac{1}{2} C^{\mu a \nu b} \Big( \partial_\mu \psi_a \partial_\nu \psi_b + 2 \partial_\mu\psi_a y^\sigma (\hat{N}_{\sigma\nu}\psi_b) + y^\sigma y^\rho (\hat{N}_{\sigma\nu} \psi_b) (\hat{N}_{\rho\mu}\psi_a) \Big) \;,
\end{equation}
\end{widetext}
where $ C^{\mu a \nu b} $ is a constant tensor constructed from various combinations of $ \delta^{\mu\nu} $ and is symmetric in $ (\mu a) \leftrightarrow (\nu b) $, and, for typographical convenience, the vector index of $ \psi $ is denoted by the Latin letter. Since the virial current is assumed to be a total derivative,
\begin{equation}
C^{\mu a \nu b}\partial_\mu\psi_a \hat{N}_{\rho\nu}\psi_b = \partial_\mu L^\mu_\rho \;,
\end{equation}
it follows that
\begin{equation} \label{KinSecond}
\frac{\delta L^\mu_\rho}{\delta\psi_a} = C^{\mu a \nu b} \hat{N}_{\rho\nu}\psi_b \;.
\end{equation}
To proceed further, an explicit form of $ L^\mu_\sigma $ should be used. For a spin-1 field, in general case, it reads
\begin{equation} \label{DerForm}
L^\mu_\sigma = \alpha \psi^2 \delta^\mu_\sigma + \beta \psi^\mu\psi_\sigma\;,
\end{equation}
where $ \alpha $ and $ \beta $ are some constants. Then, by using (\ref{KinSecond}) and (\ref{DerForm}), one can rewrite the third term in (\ref{Kinetic}) as, 
\begin{equation} \label{KinThird}
\begin{split}
\frac{1}{2} &C^{\mu a \nu b} y^\rho y^\sigma (\hat{N}_{\rho\mu}\psi_a) (\hat{N}_{\sigma\nu}\psi_b) =  \\ = \Delta\alpha y^2 &\psi^2 + \frac{\beta}{2} \left( y^2\psi^2 + (2\Delta - d)(y_\mu \psi^\mu)^2 \right)\;. 
\end{split}
\end{equation} 
Also, substituting (\ref{DerForm}) into the last term in (\ref{OnlyViaDer}) yields
\begin{equation} \label{AddSCT}
\begin{split}
\varepsilon_{\mu_0 ... \mu_d} L^{\mu_0}_\nu \omega_K^\nu \wedge \omega_P^{\mu_1} &\wedge ... \wedge \omega_P^{\mu_d} = \\ = dy^\rho \wedge \tilde{L}_\rho + (2L^\mu_\rho y^\rho y_\mu - y^2 &L^\mu_\mu)dx^1 \wedge ... \wedge dx^d \;,
\end{split}
\end{equation}
where $ \tilde{L}_\rho $ is a differential form such that $ \partial_\mu L^\mu_\rho = d \tilde{L}_\rho $. Full Lagrangian (\ref{OnlyViaDer}) is a sum of (\ref{Kinetic}) and (\ref{AddSCT}), which, as it can be explicitly verified, contains $ y^\nu(x) $ only via full derivative, $ d(y^\rho \tilde{L}_\rho) $. 

\bibliography{cftcoset}

\end{document}